\pdfoutput=1
\documentclass[11pt]{article}

\usepackage[final]{acl}

\usepackage{times}
\usepackage{latexsym}

\usepackage[T1]{fontenc}
\usepackage[utf8]{inputenc}

\usepackage{microtype}

\usepackage{inconsolata}

\usepackage{graphicx}
\usepackage{hyperref}

\title{Med-CoDE: Medical Critique based Disagreement Evaluation Framework}

\author{
 \textbf{Mohit Gupta}\textsuperscript{*},
 \textbf{Akiko Aizawa}\textsuperscript{+},
 \textbf{Rajiv Ratn Shah}\textsuperscript{*}
\\
 \textsuperscript{*}Indraprastha Institute of Information Technology Delhi, India,\\
 \textsuperscript{+}National Institute of Informatics, Tokyo, Japan
\\
 \textit{\{\href{mailto:mohit22112@iiitd.ac.in}{mohit22112}, \href{mailto:rajivratn@iiitd.ac.in}{rajivratn}\}}@iiitd.ac.in$^*$\\
 \textit{\href{mailto:aizawa@nii.ac.jp}{aizawa}}@nii.ac.jp$^+$
}

\begin{document}
\maketitle
\begin{abstract}
The emergence of large language models (LLMs) has significantly influenced numerous fields, including healthcare, by enhancing the capabilities of automated systems to process and generate human-like text. However, despite their advancements, the reliability and accuracy of LLMs in medical contexts remain critical concerns. Current evaluation methods often lack robustness and fail to provide a comprehensive assessment of LLM performance, leading to potential risks in clinical settings. In this work, we propose Med-CoDE, a specifically designed evaluation framework for medical LLMs to address these challenges. The framework leverages a critique-based approach to quantitatively measure the degree of disagreement between model-generated responses and established medical ground truths. This framework captures both accuracy and reliability in medical settings. The proposed evaluation framework aims to fill the existing gap in LLM assessment by offering a systematic method to evaluate the quality and trustworthiness of medical LLMs. Through extensive experiments and case studies, we illustrate the practicality of our framework in providing a comprehensive and reliable evaluation of medical LLMs.
\end{abstract}

\section{Introduction}
Medical Question Answering systems based on Large Language Models represent a significant leap in leveraging artificial intelligence for healthcare. These systems are designed to process and respond to medical queries. The primary aim of Medical QA LLMs is to provide accurate, reliable, and timely information to support clinicians, researchers, and patients. Evaluating the performance of these LLMs is crucial to ensure their reliability and effectiveness in real-world medical applications. Performance evaluation typically involves assessing the accuracy, relevance, and coherence of the generated responses compared to established medical standards or expert opinions.
\begin{figure}
    \centering
    \includegraphics[width=\linewidth]{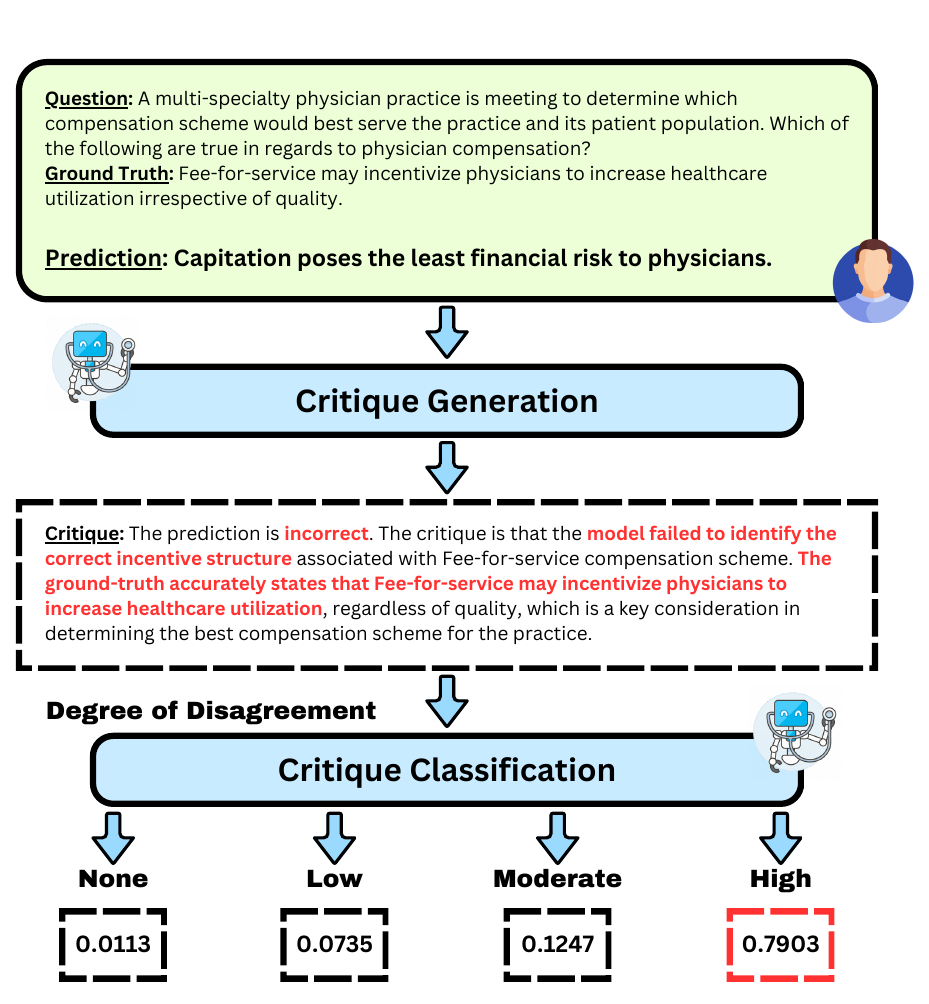}
    \caption{Med-Code Framework}
    \label{medcode-framework}
\end{figure}

Traditional methods for evaluating text generation, such as string similarity metrics (e.g., METEOR, BLEU, ROUGE), have been used widely across various domains. These metrics compare the overlap between generated and reference text-based on the n-gram matching, synonymy, and paraphrasing. While effective in general text generation tasks, these metrics pose significant limitations in the medical QA domain. Medical texts often require precise and contextually accurate responses where minor discrepancies can lead to substantial misunderstandings or clinical errors. Traditional metrics fail to capture the nuanced medical context, thereby providing an inadequate measure of LLM performance in this sensitive field.

To address the shortcomings of traditional evaluation methods, researchers have started exploring the use of LLMs themselves for evaluating other LLMs. Frameworks such as Harness~\cite{eval-harness}, DeepEval\footnote{\url{https://docs.confident-ai.com/}}, MLFlow\footnote{\url{https://mlflow.org/}} represent this shift towards LLM-assisted evaluation. These frameworks aim to provide more contextual and comprehensive evaluations by leveraging the advanced capabilities of LLMs to understand the generated responses. Despite these advancements, the current LLM-assisted evaluation methods still lack a structured approach to quantifying disagreement and assessing reliability.
\begin{figure*}[ht]
    \centering
    \includegraphics[width=\linewidth]{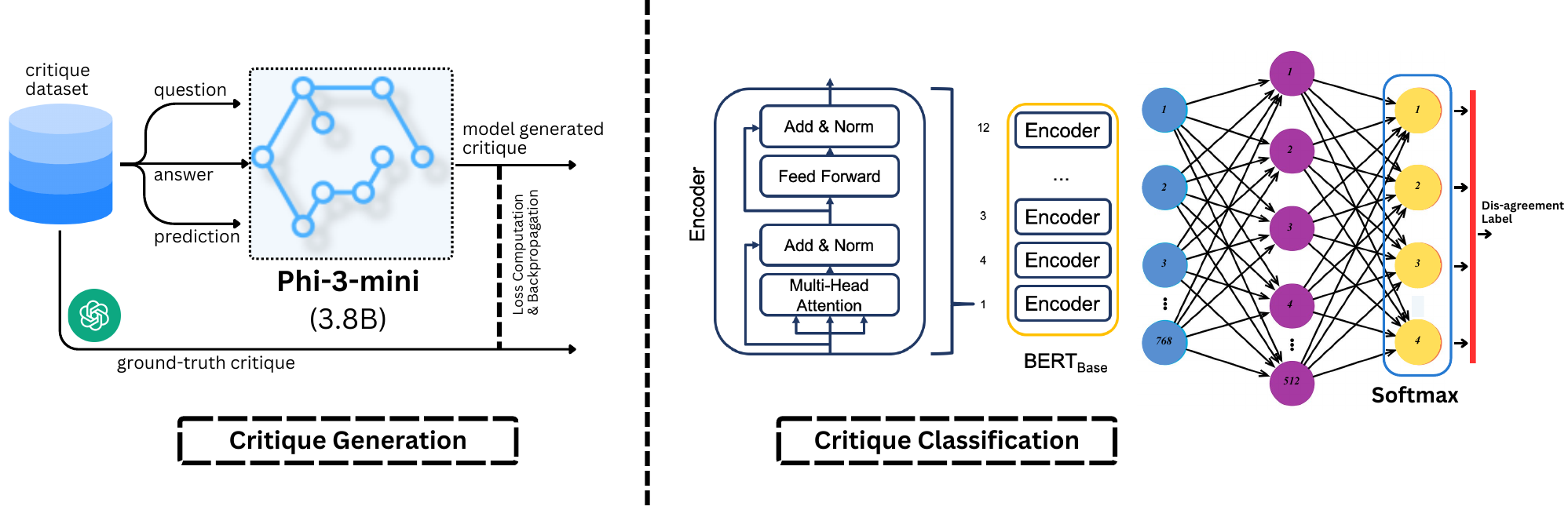}
    \caption{The overall Fine-tuning pipeline for Critique Generator \& Classifier.}
    \label{pipeline}
\end{figure*}

This research paper presents an reliable evaluation framework tailored for Medical QA LLMs. Drawing inspiration from the work of~\cite{wang2023shepherdcriticlanguagemodel}, our framework introduces a critique-based methodology that quantitatively assesses the discrepancies between model-generated responses and established medical ground truths. By employing a critique model, we analyzed the differences in LLM outputs and provide a comprehensive evaluation of their accuracy and reliability. The visual representation of Med-Code framework is shown in Fig.~\ref{medcode-framework}. 

The contributions of this work are as follows.
\begin{itemize}
    \item We curated a specialized medical critique dataset, incorporating medical Q\&A pairs from benchmark datasets such as Medqa~\cite{zhang2018medicalexamquestionanswering},  Medmcqa~\cite{pal2022medmcqalargescalemultisubject}. etc. The dataset includes responses from various medical language models (LLMs) and a degree of disagreement label between the ground-truth answers and the models' responses.
    \item We developed an advanced evaluation pipeline based on the Shepherd model~\cite{wang2023shepherdcriticlanguagemodel}, where we fine-tuned the Phi-3 model for generating critiques and employed a BERT model for classifying them.
    \item To demonstrate the effectiveness of our evaluation framework, we conducted comprehensive experiments across four medical benchmark datasets, utilizing diverse evaluation techniques to ensure robust validation.
\end{itemize}

\section{Related Work}
\label{related_work}
This section discusses related work in the field of evaluation, highlighting previous contributions. Our motivation stems from the Shepherd Model~\cite{wang2023shepherdcriticlanguagemodel}, which introduces a large language model designed to generate critiques of model responses to given prompts. We extend this work by using critiques to evaluate discrepancies between model responses and ground truth.

Recent studies have shown that traditional metrics such as METEOR~\cite{banerjee-lavie-2005-meteor}, ROUGE~\cite{lin-2004-rouge} and BLEUScore~\cite{zhang2020bertscoreevaluatingtextgeneration} are inadequate for accurately evaluating open-ended generation tasks due to their reliance on reference text~\cite{chiang-lee-2023-large,gu2020perceptionscorelearnedmetric,guan2021openmevabenchmarkevaluatingopenended,effectsofsemantics,10.1016/j.asoc.2021.107815}. Advances have led to new research using LLMs as evaluators, demonstrating their potential to overcome these limitations~\cite{kim2024prometheusinducingfinegrainedevaluation,kocmi2023largelanguagemodelsstateoftheart,liu2024xevalgeneralizablemultiaspecttext,liu2023calibratingllmbasedevaluator}. Notably, approaches employing powerful LLMs like GPT-4 have achieved remarkable performance~\cite{fu2023gptscoreevaluatedesire,liu2023gevalnlgevaluationusing}. However, current LLM-based evaluators exhibit shortcomings in robustness, as their performance is highly sensitive to prompts, leading to instability in the evaluation process. Recent studies have sought to address these challenges by generating explanations for evaluation outputs~\cite{chiang-lee-2023-large}, but this approach does not inherently improve robustness or reliability due to issues such as hallucinations~\cite{xu2023instructscoreexplainabletextgeneration}.

In the context of medical AI, where accuracy and reliability are crucial, several research efforts propose strategies to evaluate LLM responses. An automatic evaluation metric and algorithm for LLMs’ clinical capabilities is proposed in~\cite{liu2024automaticevaluationllmsclinical}, featuring a multi-agent framework with Retrieval-Augmented Evaluation (RAE) to assess the behaviors of a doctor agent.~\cite{humanely} propose a structured method for comprehensive human evaluation of LLM outputs, introducing the HumanELY guidance and tool.~\cite{liao2024automaticinteractiveevaluationlarge} introduce the Automated Interactive Evaluation (AIE) framework, which provides a dynamic, realistic platform for assessing LLMs through multi-turn doctor-patient simulations.

\section{Methodology}
\label{methodology}
In this section, we discuss the process of creating a fine-tuning dataset for the medical domain critique model, the approach we used for fine-tuning the LLM, and the development of classification model.

\subsection{Dataset}
\label{sec:dataset_curation}
In this research, we curated a specialized dataset using the OpenAI GPT-4 model to build a fine-tuning dataset for our critique generation model. Our final critique dataset comprises \textit{38,819} samples, with an average critique length of \textit{58.95} words. This dataset enables us to assess how well LLM responses align with ground-truth answers and to measure the degree of disagreement, providing a robust foundation for evaluating the performance of medical QA LLMs.

For medical domain data, we selected and combined small random subsets from standard medical QA datasets including Medqa~\cite{zhang2018medicalexamquestionanswering}, Medmcqa~\cite{pal2022medmcqalargescalemultisubject}, MMLU~\cite{hendrycks2021measuringmassivemultitasklanguage}, and Pubmedqa~\cite{jin2019pubmedqadatasetbiomedicalresearch}. These datasets encompass medical question-answer pairs from various medical fields, covering different levels of difficulty and types of questions. This comprehensive combination ensures that our critique model can effectively evaluate both objective and subjective questions.
\begin{figure}[ht]
    \centering
    \includegraphics[width=\columnwidth]{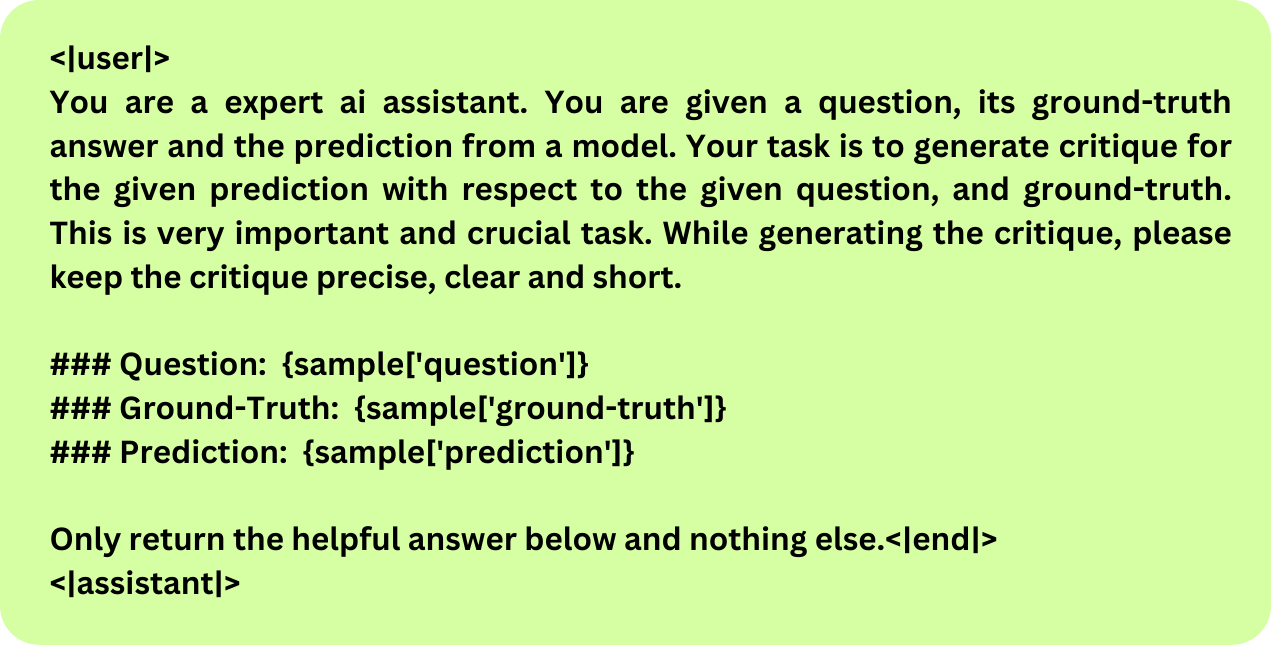}
    \caption{Critique Generation Prompt Template}
    \label{critique_generation_prompt}
\end{figure}

After merging the random subsets, we employed SOTA medical domain LLMs, such as Meditron-7B~\cite{chen2023meditron70bscalingmedicalpretraining}, SelfBioRAG-7B~\cite{jeong2024improvingmedicalreasoningretrieval}, to generate answers for each question. Each response was then critically evaluated using OpenAI GPT-4, which assigned a disagreement label from one of four categories: None, Low, Medium, and High. A High disagreement label indicates that the model-generated response is entirely incorrect and does not align with the ground truth in any aspect, whereas a None disagreement label signifies that the response is accurate and fully aligns with the ground truth without any extraneous information. In Low disagreement label the response is mostly accurate with minor additional details or slight deviations from the ground truth, lastly, the Moderate disagreement label, the response contains a mix of correct and incorrect information, with significant deviations from the ground truth, meaning the model is hallucinating.

\subsection{Models}
\label{sec:models}
To build this lightweight evaluation framework, we employed two small models: \textbf{\textit{Phi-3 3.8B}}~\cite{abdin2024phi3technicalreporthighly} for generating critiques \& \textbf{\textit{BERT}}~\cite{devlin2019bertpretrainingdeepbidirectional} for classifying the critiques. Although larger models with superior text generation capabilities and understanding are available, our objective was to create a domain-specific model tailored for a single task. Hence, these models were chosen. The visual representation of fine-tuning model architectures is shown in Fig.~\ref{pipeline}.  This integrated pipeline proved efficient across all aspects, including computation, speed, and accuracy.

\section{Experiments}
\label{experiments}
In this section, we will delve into the experiment setup we have used for building this framework. It is divided into two subsections, first is for the critique generation model, and second is for the critique classification model.

\subsection{Critique Generation Model}
The objective of this model is to generate critiques based on a given question, its ground-truth answer, and the model's response. For this purpose, we employed the phi-3-mini model, which contains 3.8 billion parameters.

The hyperparameters configured for fine-tuning include \textit{5} epochs, a batch size of \textit{128}, a learning rate of \textit{1.41e-5}, and the AdamW 8-bit optimizer. We utilized the LORA technique for efficient fine-tuning, with a rank parameter \( r = 16 \). The training process consumed an average of \textit{20} GBs VRAM and required approximately \textit{4-5} hours of GPU time. The data set was split into \textit{30,000} samples for training, \textit{4,409} for testing, and \textit{4,410} for validation. The prompt template used in the fine-tuning and inference is given in Fig.~\ref{critique_generation_prompt}. 

\begin{figure*}[ht]
    \centering
    \includegraphics[width=\linewidth]{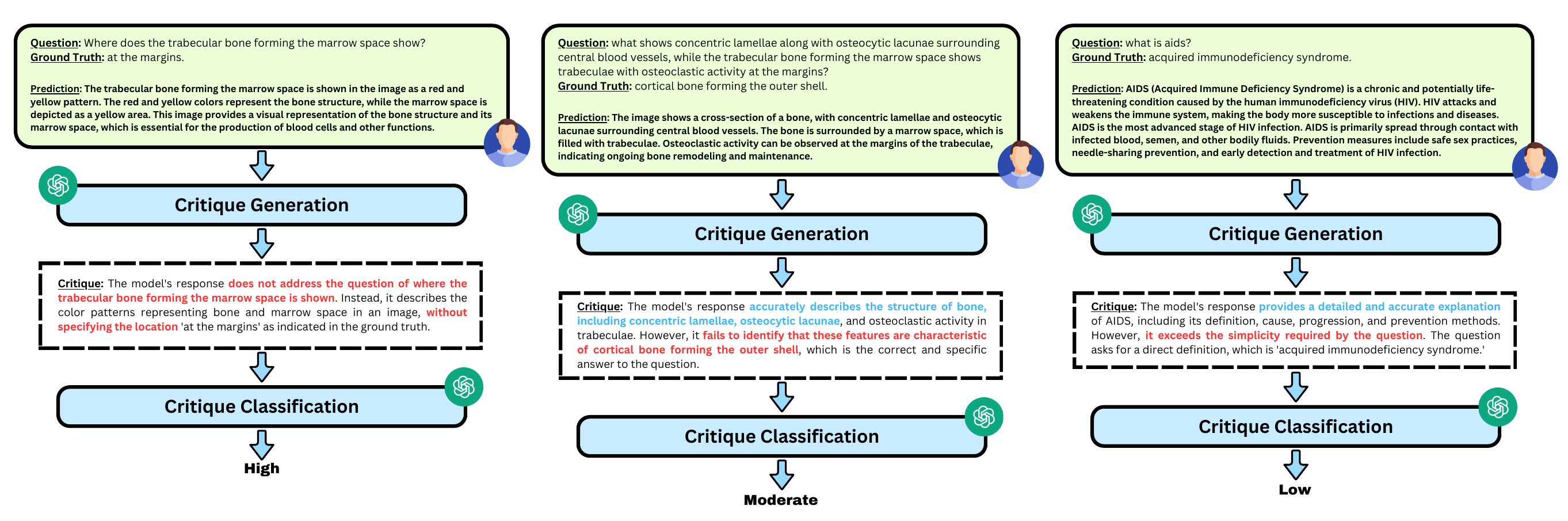}
    \caption{Critique data samples with different dis-agreement Labels}
    \label{fig:dataset_examples}
\end{figure*}

Examples for each class of disagreement are provided in Fig.~\ref{fig:dataset_examples}. These examples illustrate that the critiques generated by the model are highly precise and clear in identifying discrepancies between the ground-truth answers and the model's predictions, thereby supporting the efficacy of the fine-tuning process. To evaluate the quality of the dataset, we conducted a quality assessment on a small subset, as detailed in Section \ref{quality_assessment}.

\subsection{Critique Classification Model}
For the critique classification model, we utilized the BERT base model, which contains \textit{110M} parameters. This model is lightweight yet offers a deep bidirectional understanding of context, effectively capturing nuanced language patterns. The architecture of the entire classification network is depicted in Fig.~\ref{pipeline}.
\begin{table}[ht]
  \centering
  \renewcommand{\arraystretch}{1.1}
  \setlength{\tabcolsep}{4\tabcolsep}
  \begin{tabular}{|l|r|}
    \hline
    \textbf{Framework} & \textbf{Accuracy} \\
    \hline
    \textbf{GPT-3.5} & 78.12 \\\hline
    \textbf{Med-Code} & 71.72 \\
    \hline
  \end{tabular}
  \caption{Human Evaluation Results of Disagreement Classification}
  \label{human_evaluation}
\end{table}

The hyperparameters configured for fine-tuning are \textit{25} epochs, a learning rate of \textit{1e-3}, a dropout rate of \textit{0.3}, a batch size of \textit{16}, and a maximum sequence length of \textit{208} tokens. The fine-tuning process employed a weighted average of all classes, with class weights specified as [\textit{5.96, 1.34, 0.83, 0.52}]. The divergence function used is the Negative Log Likelihood. The total GPU utilization for fine-tuning this network is \textit{2,771} MiB with \textit{1} hour of GPU time. The data split used in this model training is \textit{27,173} samples for training, \textit{5,823} samples for validation, and \textit{5,823} samples for testing. 

We conducted a performance analysis of OpenAI’s GPT-3.5 and our proposed framework, Med-Code, on a human labeled subset of 265 randomly selected samples. Each model received a question, a ground-truth answer, and the model’s prediction, and we evaluated their accuracy in disagreement classification based on the critiques they generated. As shown in Table.~\ref{human_evaluation}, GPT-3.5 correctly classified approximately 207 out of 265 samples, and our proposed Med-Code framework produced results comparable to those of GPT-3.5 which is around 190 samples. 

\section{Results \& Analysis}
\label{results}
To assess the effectiveness of evaluating responses from large language models, we conducted experiments on four medical benchmark datasets: Medqa~\cite{zhang2018medicalexamquestionanswering}, Medmcqa~\cite{pal2022medmcqalargescalemultisubject}, Pubmedqa~\cite{jin2019pubmedqadatasetbiomedicalresearch}, and Mmlu~\cite{hendrycks2021measuringmassivemultitasklanguage}. These datasets are widely used in medical benchmarking and consist of objective-type questions. Our analysis focused on the test sets of these datasets using three LLMs: LLaMA-3~\cite{llama3modelcard}, Mistral~\cite{jiang2023mistral7b}, and BioMistral~\cite{labrak-etal-2024-biomistral}. We selected these LLMs due to their demonstrated superior performance on general tasks and medical benchmarks.
\begin{table*}[ht]
  \centering
  \renewcommand{\arraystretch}{1.2}
  \setlength{\tabcolsep}{\tabcolsep}
  \begin{tabular}{|l|r|r|r|r|r|r|r|r|r|r|r|r|r|}
    \hline
    \textbf{Dataset}& \multicolumn{2}{c|}{\textbf{Automatic Evaluation}} & \textbf{LLM-Accuracy} & \multicolumn{4}{c|}{\textbf{Dis-agreement Evaluation}}\\
    \hline
    & \small \textbf{Meteor} & \small \textbf{Rouge-L} & & \small \textbf{None} $\uparrow\uparrow$ & \small \textbf{Low} $\uparrow$ & \small \textbf{Moderate} $\downarrow$ & \small \textbf{High} $\downarrow\downarrow$ \\\hline
    \multicolumn{8}{|c|}{\textbf{Results for LLaMA-3}} \\\hline
    \small \textbf{MEDQA USMLE} & 0.51 & 0.52 & 0.69 & 0.53 & 0.22 & 0.13 & 0.12 \\\hline
    \small \textbf{MEDMCQA} & 0.12 & 0.26 & 0.53 & 0.47 & 0.32 & 0.13 & 0.07 \\\hline
    \small \textbf{PUBMEDQA} & 0.11 & 0.12 & 0.39 & 0.55 & 0.30 & 0.10 & 0.05 \\\hline
    \small \textbf{MMLU} & 0.71 & 0.71 & 0.70 & 0.57 & 0.31 & 0.09 & 0.04 \\
    \hline
    \multicolumn{8}{|c|}{\textbf{Results for BioMistral 7B}} \\\hline
    \small \textbf{MEDQA USMLE} & 0.14 & 0.07 & 0.74 & 0.44 & 0.29 & 0.16 & 0.11 \\\hline
    \small \textbf{MEDMCQA} & 0.16 & 0.08 & 0.61 & 0.35 & 0.39 & 0.18 & 0.08 \\\hline
    \small \textbf{PUBMEDQA} & 0.21 & 0.16 & 0.73 & 0.54 & 0.30 & 0.11 & 0.05 \\\hline
    \small \textbf{MMLU} & 0.33 & 0.19 & 0.70 & 0.32 & 0.41 & 0.19 & 0.07 \\
    \hline
    \multicolumn{8}{|c|}{\textbf{Results for Mistral 7B v2.0}} \\\hline
    \small \textbf{MEDQA USMLE} & 0.16 & 0.12 & 0.68 & 0.47 & 0.28 & 0.15 & 0.01 \\\hline
    \small \textbf{MEDMCQA} & 0.56 & 0.11 & 0.56 & 0.33 & 0.38 & 0.20 & 0.08 \\\hline
    \small \textbf{PUBMEDQA} & 0.21 & 0.19 & 0.68 & 0.60 & 0.26 & 0.09 & 0.05 \\\hline
    \small \textbf{MMLU} & 0.37 & 0.25 & 0.65 & 0.36 & 0.37 & 0.19 & 0.07 \\
    \hline
  \end{tabular}
  \vspace{0.3cm}
  \caption{Evaluation Results for LLaMA-3, BioMistral 7B and Mistral 7B v2.0}
  \label{tab:merged}
\end{table*}

We utilized Meteor and Rouge-L scores for automatic evaluation, the LLaMA-3 model for LLM-assisted evaluation, and our Med-Code framework to analyze LLM performance comprehensively. Med-Code categorizes responses into four degrees of disagreement, where an ideal model would show the highest average probability for “None” disagreement and the lowest for “High” disagreement. Detailed descriptions of each disagreement label are provided in the Section~\ref{sec:dataset_curation}.

In Table~\ref{tab:merged}, LLaMA-3, BioMistral, and Mistral models were used for inference. LLaMA-3 performed best on the MMLU dataset, achieving high scores in both automatic and LLM-assisted evaluations. Med-Code results showed that the “None” disagreement probability was the highest, indicating strong alignment between the model’s responses and the ground-truth answers. Conversely, the “High” disagreement probability was the lowest, supporting the model’s accuracy.

The automatic evaluation results for BioMistral, a medical domain-specific LLM, did not convey significant information due to its poor string/semantic matching. However, BioMistral outperformed Mistral in LLM-assisted evaluation accuracy across all datasets, which was expected. 

There was a strong positive correlation between accuracy and “None” disagreement probability, demonstrating that Med-Code effectively identified correct responses. Additionally, there is a positive correlation between METEOR scores and a 'Low' disagreement probability, suggesting that the low semantic relation between ground truth and model predictions. The low positive correlation between LLM-assisted accuracy and both 'Moderate' and 'High' disagreement probabilities confirmed instances where the models hallucinated or produced incorrect results. 

When examining the correlation between automatic evaluation scores like METEOR and ROUGE-L scores and LLM accuracy, the correlation is inconsistent across different LLMs. This inconsistency may be due to the fact that automatic metrics are based on string matching, while LLM-assisted accuracy relies on the model's knowledge and logic. For example,

\vspace{0.1cm}
\noindent\colorbox{blue!5}{\parbox{\linewidth}{\textit{"If the model generates medicine \textbf{$X$} for disease \textbf{$D$}, but the ground truth answer lists medicine \textbf{$Y$} for the same disease, the automatic evaluation scores might be low. However, the LLM-assisted accuracy could still be correct because the model knows that $X$ is equivalent to $Y$ for disease \textbf{$D$}."}}}

\subsection{Human Evaluation}
\label{quality_assessment}
To assess the quality of the critique data generated by the OpenAI model for fine-tuning purposes, we conducted a thorough evaluation on a randomly selected subset of 265 samples. Each sample was manually reviewed to determine how effectively the model understood the relationship between the ground-truth answer and the model's prediction, and whether it could accurately identify minute discrepancies and details within the predictions.

Upon analysis, we found that approximately 240 out of the 265 samples (about \textit{91\%}) were accurately critiqued. The generated critiques successfully highlighted the flaws and discrepancies between the ground-truth and the predictions, demonstrating the model's capability to provide precise and detailed feedback. This quality assessment validates the reliability of the generated data for fine-tuning the critique generation model. The ground-truth critiques are noted for their clarity and precision, effectively pinpointing subtle differences between the ground-truth answers and the model's predictions. This ensures that the data can be effectively used for fine-tuning the critique generation model, allowing it to learn and adapt with high accuracy and precision.

\section{Conclusion}
\label{conclusion}
In this work, we introduce Med-CoDE, an evaluation framework designed to assess the performance of Medical LLMs using critiques and degrees of disagreement. Med-CoDE excels in identifying subtle discrepancies between ground-truth answers and model predictions, offering a nuanced evaluation with four levels of disagreement. These levels provide insights into the model's behavior, such as hallucinations, accuracy, and adherence to the question. Our framework aids researchers in pinpointing areas where LLMs fall short, enabling targeted improvements. Extensive experiments on standard medical benchmark datasets demonstrate Med-CoDE's effectiveness in thoroughly and efficiently analyzing model behavior. This robust evaluation method is crucial for advancing the reliability and safety of AI-driven healthcare solutions. This evaluation framework is adaptable for assessing large language models across various domain-specific tasks as well as general tasks, simply by modifying the critique dataset.

\section{Limitations}
\label{limitations}
In this paper, we assess both automatic and human evaluation. Despite experimenting with a substantial number of data examples and utilizing human annotators to the best of our financial capabilities, there is room for further enhancement. Limited access to the costly OpenAI APIs meant that we used these resources judiciously, focusing on crucial areas. Additionally, computational constraints restricted the scope of our experiments. Nonetheless, these limitations highlight opportunities for future work to expand and refine the proposed framework with more extensive experimental analysis and resource allocation.

\section{Ethical Considerations}
\label{ethical_considerations}
The Med-CoDE framework, designed to assess the reliability and accuracy of medical LLMs, operates within a domain where the potential consequences of errors are particularly significant, given the direct impact on patient care and treatment outcomes.

In this work, only the publicly available standard benchmark medical QA datasets are used for training and evaluations. The Med-CoDE framework aims to enhance the evaluation of LLMs to ensure they meet rigorous standards of accuracy and reliability. However, it is essential to recognize that even well-evaluated models are not infallible and should not replace human judgment. Instead, they should be used as tools to support healthcare professionals, who must remain the final arbiters in clinical decision-making.

By addressing these ethical considerations, the Med-CoDE framework can contribute to the responsible development and deployment of medical LLMs, ultimately supporting safer and more effective healthcare solutions.

\section{Acknowledgments}
This work was supported by Cross-ministerial Strategic Innovation Promotion Program (SIP) on “Integrated Health Care System” Grant Number JPJ012425.

% Bibliography entries for the entire Anthology, followed by custom entries
%\bibliography{anthology,custom}
% Custom bibliography entries only
% \bibliography{software}

\end{document}